# Second-order effects of the magnetic vector potential in the Aharonov-Bohm experiment

Keith J. Kasunic

*Optical Systems Group LLC, Tucson, AZ 85728 USA*
*Dept. of Physics & Optical Science, Univ. of North Carolina, Charlotte, NC 28223*

## Abstract

Recent experiments with the Aharonov-Bohm geometry have shown that, in addition to an electron-interference fringe shift, there is also a lateral displacement of the electron diffraction envelope. In this paper, we derive a displacement force based on a second-order expansion of the magnetic vector potential. The analysis illustrates the conservation of canonical angular momentum, where the mechanical angular momentum and field angular momentum sum to a constant of the motion; the azimuthal force required to change the mechanical momentum is thus supplied by changes in field momentum associated with the second-order vector potential term. Our results are consistent with all known Aharonov-Bohm experiments, including interference fringe shifts, lateral displacement forces, and the absence of longitudinal forces.

## I. BACKGROUND

Recent experiments with the Aharonov-Bohm (AB) geometry have shown that, in addition to an electron-interference fringe shift, there is also a lateral displacement of the electron diffraction envelope [1]. This raises yet again the question of what "forces" are involved with such displacements in regions of space where there are neither electric nor magnetic fields. Possibilities previously considered in the literature have focused on quantum interpretations such as Bohm's concept of the quantum force [2], the collimated-beam scattering analyses of Shelankov [3] and Berry [4], and the mathematical derivations of Keating and Robbins [5]. In this paper, we propose a displacement force based on a second-order expansion of the magnetic vector potential. This mechanism does not require a penetration of the electron's electromagnetic field into the solenoid; our results are consistent with all known AB experiments, including lateral forces [1], interference fringe shifts [6]-[7], and the absence of longitudinal forces [8].

In the derivation of the well-known Lorentz force law, the magnetic force on a moving charged particle is obtained using a first-order Taylor series expansion of the magnetic vector potential $\mathbf{A}(x,y,t)$. This first-order expansion for changes in $\mathbf{A}(x,y,t)$ gives the total (or "material" or "substantial" or "hydrodynamic") time derivative $d\mathbf{A}/dt$ in the Lagrangian description of a reference frame moving with the particle [9]-[10]

$$\frac{d\mathbf{A}(x,y,t)}{dt} \approx \frac{\partial \mathbf{A}}{\partial t} + v_x \frac{\partial \mathbf{A}}{\partial x} + v_y \frac{\partial \mathbf{A}}{\partial y} = \frac{\partial \mathbf{A}}{\partial t} + (\mathbf{v} \cdot \nabla)\mathbf{A} \qquad (1)$$

where $v_x = \dot{x}$ and $v_y = \dot{y}$ are the group velocity components, giving gradient terms that are known as "convective" changes in $\mathbf{A}$ along the path of motion. Physically, the $(\mathbf{v} \cdot \nabla)\mathbf{A}$ term represents point-to-point changes in the velocity (i.e., an acceleration) of a particle as it moves through first-order spatial variations in the vector potential of the form $\partial \mathbf{A}/\partial x$. Including both electric and magnetic fields, the electromagnetic force $\mathbf{F}_{EM}$ is obtained using a Lagrangian analysis [9]

$$\mathbf{F}_{EM} = e(\mathbf{E} + \mathbf{v} \times \mathbf{B}) \qquad (2)$$

for a particle with charge $e$ moving with a non-relativistic velocity $\mathbf{v}(x,y,t)$ through an electric field $\mathbf{E} = -\partial \mathbf{A}/\partial t$ and a magnetic field $\mathbf{B} = \nabla \times \mathbf{A}$.

## II. SECOND-ORDER VECTOR POTENTIAL

In this paper, we incorporate point-to-point changes in acceleration (i.e., the "jolt" or "jerk", $\dddot{\mathbf{x}}$) [11] with a second-order expansion of the vector potential. Using operator notation, the change in $\mathbf{A}(x,y,t)$ using a second-order Taylor-series expansion in a Lagrangian frame moving with the particle is given by

$$\Delta \mathbf{A}(x,y,t) \approx \left[\Delta x \frac{\partial}{\partial x} + \Delta y \frac{\partial}{\partial y} + \Delta t \frac{\partial}{\partial t}\right] \mathbf{A}(x,y,t) + \frac{1}{2!}\left[\Delta x \frac{\partial}{\partial x} + \Delta y \frac{\partial}{\partial y} + \Delta t \frac{\partial}{\partial t}\right]^2 \mathbf{A}(x,y,t) + \ldots \qquad (3)$$

Ignoring the first-order terms and using stationary (steady-state) solutions with $\mathbf{E} = -\partial \mathbf{A}/\partial t = 0$ and a velocity profile $\mathbf{v}(x,y)$ which varies from point to point but does not change over time at any fixed point [12], we obtain the second-order total derivative $d^2\mathbf{A}/dt^2$

$$\frac{d^2 \mathbf{A}[x(t),y(t)]}{dt^2} \approx \frac{1}{2}\left[v_x^2 \frac{\partial^2 \mathbf{A}}{\partial x^2} + 2v_x v_y \frac{\partial^2 \mathbf{A}}{\partial x \partial y} + v_y^2 \frac{\partial^2 \mathbf{A}}{\partial y^2}\right] = \frac{1}{2}\left[v_x \frac{\partial}{\partial x} + v_y \frac{\partial}{\partial y}\right]^2 \mathbf{A} \qquad (4)$$

where the non-linear term on the right-hand side is given by $(\mathbf{v} \cdot \nabla)^2 \mathbf{A}$. Physically, this non-linear term represents convective changes in acceleration as a particle moves from point-to-point through second-order spatial variations in the vector potential of the form $\partial^2 \mathbf{A}/\partial x^2$.

Bringing the first-order terms back into the equation, the second-order term in Eq. (4) must be integrated over time to obtain its contribution to the total derivative

$$\frac{d\mathbf{A}[x(t), y(t)]}{dt} \approx v_x \frac{\partial \mathbf{A}}{\partial x} + v_y \frac{\partial \mathbf{A}}{\partial y} + \int \frac{d^2 \mathbf{A}}{dt^2} dt = (\mathbf{v} \cdot \nabla)\mathbf{A} + \frac{1}{2} \int (\mathbf{v} \cdot \nabla)^2 \mathbf{A} \, dt \quad (5)$$

from which a Lagrangian analysis then identifies a force

$$m\ddot{\mathbf{x}} = e\left[ (\mathbf{v} \times \mathbf{B}) - \frac{1}{2} \int (\mathbf{v} \cdot \nabla)^2 \mathbf{A} \, dt \right] \quad (6)$$

that includes both magnetic field and convective second-order terms. It is this equation that we use in the next section to analyze the AB effect with $\mathbf{B} = 0$.

III. AHARONOV-BOHM EFFECT

Figure 1 shows a physical demonstration of the AB experiment, where electron-interference fringe positions are shifted based on the presence of a non-local magnetic field [13]. To date, the idea of a "force" in interpreting this fringe shift – electromagnetic [14], quantum [1-4, 15], or non-existent [16] – has been problematic.

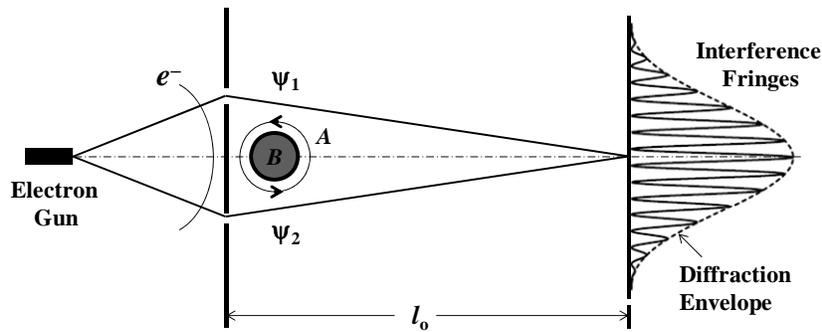

Fig. 1 – Schematic of the Aharonov-Bohm (AB) two-slit interference experiment, where the magnetic field **B** is zero in the region outside the solenoid, yet the matter waves $\psi_1$ and $\psi_2$ for the electron $e^-$ recombine with different maxima and minima locations, depending on the vector potential **A**.

In a recent paper, we showed that the magnetic vector potential in the Aharonov-Bohm effect acts as a quantum "phase plate", changing in a non-dispersive, gauge-invariant manner the phase difference between $\psi_1$ and $\psi_2$ on the upper and lower halves of the solenoid [17]. Physically, this was shown to be a direct result of the vector potential adding (or subtracting) field momentum $e\mathbf{A}$ to (or from) the initial electron momentum $\mathbf{p}_o$

$$\mathbf{p} = \mathbf{p}_o - e\mathbf{A} \qquad (7)$$

The difference in momentum changes the de Broglie wavelength of these two waves

$$\lambda(\mathbf{A}) = \frac{h}{|\mathbf{p}_o - e\mathbf{A}|} \qquad (8)$$

thus producing a phase shift with a resulting change in fringe position when $\psi_1$ and $\psi_2$ are interfered. This change in fringe position is analogous to the principle behind phased-array radar for non-mechanical beam steering [18], except in this case we are of course dealing with electron matter waves, rather than the electromagnetic waves used in radar and ladar.

In this paper, we see from Eq. (6) a way to introduce a force in the AB effect. In this section, we estimate this force based on the electron moving through the second-order radial variations of the solenoid's vector potential. To do so, we use a 2D Lagrangian analysis in polar ($r$-$\theta$) coordinates to clearly identify the radial and azimuthal terms contributing to the force.

The Lagrangian using polar coordinates is given by

$$L(r, \dot{r}, \dot{\theta}) = T - V = \frac{1}{2}m(\dot{r}^2 + r^2\dot{\theta}^2) + e(\dot{r}A_r + r\dot{\theta}A_\theta) \qquad (9)$$

for a velocity $\mathbf{v} = v_r\hat{\mathbf{r}} + v_\theta\hat{\boldsymbol{\theta}} = \dot{r}\hat{\mathbf{r}} + r\dot{\theta}\hat{\boldsymbol{\theta}}$ and the potential energy $V$ of a charged particle in a magnetic vector potential $\mathbf{A}$ given by $-e\mathbf{v}\cdot\mathbf{A}$. It is also convenient to use the Coulomb gauge ($\nabla \cdot \mathbf{A} = 0$), in which case $A_r = 0$. There is thus only a tangential component $A_\theta(r)$, found from the circulation integral on the left-hand side of Eq. (10) and the use of Stokes' Theorem, giving

$$\oint \mathbf{A} \cdot d\mathbf{s} = \iint (\nabla \times \mathbf{A}) \cdot d\mathbf{S} = \iint \mathbf{B} \cdot d\mathbf{S} \equiv \Phi_B \quad (10)$$

outside the solenoid for a path length d**s**, an enclosed cross-sectional area d**S**, a magnetic field **B** inside the solenoid, and a magnetic flux $\Phi_B$ encircled by the closed path. For a circular path around the solenoid, we find that $A_\theta(r) = \Phi_B/2\pi r$ for $r \geq$ the solenoid radius $R$.

Using the polar-coordinate Lagrangian and the Coulomb gauge, we have an equation of motion for the $r$-coordinate

$$\frac{d}{dt}\left(\frac{\partial L}{\partial \dot{r}}\right) = \frac{\partial L}{\partial r} \rightarrow m(\ddot{r} - r\dot{\theta}^2) = e\dot{\theta}\left[A_\theta(r) + r\frac{\partial A_\theta(r)}{\partial r}\right] \quad (11)$$

and a separate equation for the $\theta$-coordinate

$$\frac{d}{dt}\left(\frac{\partial L}{\partial \dot{\theta}}\right) = \frac{\partial L}{\partial \theta} \rightarrow \frac{d}{dt}[mr^2\dot{\theta} + erA_\theta(r)] = 0 \quad (12)$$

where it appears that the well-known expression [19] for the radial acceleration $a_r = \ddot{r} - r\dot{\theta}^2$ is not zero. This is not correct, however, as the $r\partial A_\theta(r)/\partial r$ term on the right-hand side (RHS) of Eq. (11) has the same magnitude but opposite sign as $A_\theta(r)$, thus reducing the RHS to zero and giving $a_r = 0$.

It is in the azimuthal equation where the AB force due to the second-order vector-potential expansion is to be found. Before getting to that, we first notice an application of Noether's Theorem, where the quantity in brackets in Eq. (12) is conserved, given that its time-derivative is equal to zero. This quantity is the *canonical angular momentum*; it consists of a mechanical (or "kinetic") angular momentum term $mr^2\dot{\theta}$ and a field angular momentum term $erA_\theta(r)$, whose sum is a constant of the motion.

With the results of the Lagrangian analysis depending on the total derivative d**A**/d*t* in Eq. (5), we expand the time derivative in Eq. (12), giving

$$\frac{1}{r}\frac{d}{dt}\left[mr^2\dot{\theta}+erA_\theta(r)\right]=m(r\ddot{\theta}+2\dot{r}\dot{\theta})+e\left[\frac{dA_\theta(r)}{dt}+\frac{\dot{r}}{r}A_\theta(r)\right]=0 \quad (13)$$

where the first two terms of the derivative contain the well-known expression [19] for the angular acceleration $a_\theta = r\ddot{\theta}+2\dot{r}\dot{\theta}$. Using only the first-order expansion of d$A_\theta$/d*t* from Eq. (5), we have for stationary solutions

$$\frac{dA_\theta}{dt}=(\mathbf{v}\cdot\nabla)A_\theta=v_r\frac{\partial A_\theta}{\partial r}+\frac{v_\theta}{r}\frac{\partial A_\theta}{\partial \theta}=\dot{r}\frac{\partial A_\theta}{\partial r}=-\dot{r}\frac{\Phi_B}{2\pi r^2}=-\frac{\dot{r}}{r}A_\theta(r) \quad (14)$$

and the terms in brackets on the RHS of Eq. (13) thus cancel, leading to $a_\theta = 0$. To first order, both radial and azimuthal acceleration terms are thus zero in the absence of electric and magnetic fields, fully consistent with the Lorentz force expression of Eq. (2).

We now include the second-order term from Eq. (5) in Eq. (13), resulting in

$$ma_\theta=-\frac{e}{2}\int(\mathbf{v}\cdot\nabla)^2 A_\theta(r)dt=-\frac{e}{2}\int\left(v_r\frac{\partial}{\partial r}+\frac{v_\theta}{r}\frac{\partial}{\partial \theta}\right)^2 A_\theta(r)dt=-\frac{e}{2}\int v_r^2\frac{\partial^2 A_\theta(r)}{\partial r^2}dt \quad (15)$$

as the mechanical angular force $F_\theta = ma_\theta$ on the charged particle. Note that we have not modified either the Lagrangian in Eq. (9) or the equation of motion given by Eq. (12), thus retaining the conservation of canonical angular momentum in this expression for the force. We next convert Eq. (15) to a space integral using $dr = v_r dt$ to obtain

$$F_\theta=-\frac{e}{2}\int v_r\frac{\partial^2 A_\theta(r)}{\partial r^2}dr \quad (16)$$

 

as the gauge-invariant azimuthal force for the AB effect. To evaluate this equation, we need an expression for the radial velocity distribution $v_r(r,\theta)$. We obtain this expression in Sec. IV.

## IV. POTENTIAL FLOW ANALYSIS

In this section, we take a closer look at how the velocity distribution varies with radius and polar angle. This allows us to decompose the azimuthal force into Cartesian coordinates to determine its longitudinal and transverse components, thus providing a basis for comparing our second-order model of electromagnetic forces with published Aharonov-Bohm experiments.

To determine the radial velocity distribution $v_r(r,\theta)$, we use complex variables methods for the solution of Laplace's equation $\nabla^2 \phi_A = 0$ in electrostatics, hydrodynamics, etc. [10, 20]. That is, given that $\mathbf{B} = \nabla \times \mathbf{A} = \nabla \times \nabla \phi_A = 0$ outside the solenoid (i.e., $\mathbf{A}$ is "irrotational"), we can obtain the vector potential $\mathbf{A} = \nabla \phi_A$ from a magnetic scalar potential $\phi_A(r,\theta)$. In addition, the Coulomb gauge for a solenoidal $\mathbf{A}$ requires that $\nabla \cdot \mathbf{A} = \nabla \cdot \nabla \phi_A = 0$, thus giving us Laplace's equation.

For the AB effect, we can also obtain potential-flow solutions for a velocity potential $\phi_v(r,\theta)$. To understand why, we note that with the initial electron momentum $\mathbf{p}_o = mv_o\hat{\mathbf{i}}$, the initial velocity field has neither divergence nor curl. Using Eq. (7) and the Coulomb gauge for $\mathbf{A}$, we may therefore write

$$m\nabla \cdot \mathbf{v} = -e\nabla \cdot \mathbf{A} = 0 \tag{17}$$

and

$$m\nabla \times \mathbf{v} = -e\nabla \times \mathbf{A} = 0 \tag{18}$$

showing that the general velocity field $\mathbf{v}(r,\theta) = \nabla \phi_v$ outside the solenoid is also irrotational and solenoidal, and can thus also be found from a solution to Laplace's equation. For flow with a clockwise velocity circulation $\Gamma_v$ around a cylinder of radius $R$, the velocity potential, $\phi_v(r,\theta)$ is given by [10, 20]

$$\phi_v(r,\theta) = v_o \cos\theta \left( r + \frac{R^2}{r} \right) + \frac{\Gamma_v}{2\pi}\theta \tag{19}$$

from which we obtain

$$v_r(r,\theta) = \frac{\partial \phi_v}{\partial r} = v_o \cos\theta \left( 1 - \frac{R^2}{r^2} \right) \tag{20}$$

and

$$v_\theta(r,\theta) = \frac{1}{r}\frac{\partial \phi_v}{\partial \theta} = -v_o \sin\theta \left( 1 + \frac{R^2}{r^2} \right) + \frac{\Gamma_v}{2\pi r} \tag{21}$$

for the radial and tangential velocity components $v_r$ and $v_\theta$, for $r \geq R$ and an angle $\theta$ measured in the conventional sense as counter-clockwise around the solenoid origin with respect to the positive x-axis. Note that $v_r = 0$ at $r = R$ for any azimuthal angle $\theta$, as required by the kinematic ("impenetrability") boundary condition for the radial velocity at the solenoid surface. Also note that the second term on the right-hand side of Eq. (21) has an almost negligible effect on the tangential velocity – on the order of 1 part in $10^6$ for $v_o \approx 0.6 \times 10^8$ m/s – but is included here to later illustrate the effects of the vector potential on the circulation $\Gamma_v$ and on the phase of the electron, as discussed in Ref. 17. Again using Eq. (7) with $\mathbf{p}_o = mv_o\hat{\mathbf{i}}$, we have

$$\Gamma_v \equiv \oint \mathbf{v}\cdot d\mathbf{s} = -\frac{e}{m}\oint \mathbf{A}\cdot d\mathbf{s} = -\frac{e}{m}\iint (\nabla \times \mathbf{A})\cdot d\mathbf{S} = -\frac{e}{m}\Phi_B \tag{22}$$

giving a non-zero value, as the region where $\psi_1$ and $\psi_2$ propagate across the cylindrical solenoid is not simply connected [21, 22].

As it is only the radial component of the velocity which contributes to the azimuthal acceleration $a_\theta$, we substitute Eq. (20) in the integral in Eq. (16), giving

$$F_\theta(r) = -\frac{e}{2}\int_R^\infty v_r \frac{\partial^2 A_\theta(r)}{\partial r^2} dr = -\frac{e}{2} v_o \cos\theta \int_R^\infty \left(1 - \frac{R^2}{r^2}\right) \frac{\Phi_B}{\pi r^3} dr = -v_o \cos\theta \frac{e\Phi_B}{8\pi R^2} \quad (23)$$

which we can decompose into axial and transverse components for a given set of experimental conditions for $v_o$, $R$, and $\Phi_B$.

To estimate the magnitude of this force, we use the magnetic flux required for a phase shift $\Delta\phi_{AB} = 2\pi$ in the AB effect, giving $\Phi_B = h/e = 4.135\times10^{-15}$ Wb; we also use a solenoid with a radius $R = 5$ μm to estimate the force at $r = R$ and $\theta = \pi$. The velocity is based on an initial electron energy $E_o = eV_o = 1.602 \times 10^{-15}$ J ($V_o = 10$ kV), giving a non-relativistic $v_r \approx v_o = (2E_o/m_e)^{1/2} \approx 0.6 \times 10^8$ m/s. Substituting these values in Eq. (23), we find the azimuthal force $F_\theta \approx 6.3 \times 10^{-17}$ N as the electron travels at $v_r$ through the radial gradients of the vector potential.

To obtain the Cartesian components of $F_\theta$, we decompose Eq. (23) using the coordinate transformations [22]

$$F_x = F_r \cos\theta - F_\theta \sin\theta = -F_\theta \sin\theta = v_o \sin\theta \cos\theta \frac{e\Phi_B}{8\pi R^2} \quad (24)$$

and

$$F_y = F_r \sin\theta + F_\theta \cos\theta = F_\theta \cos\theta = -v_o \cos^2\theta \frac{e\Phi_B}{8\pi R^2} \quad (25)$$

for which the longitudinal force $F_x$ in the direction of propagation – with a $\cos\theta\cdot\sin\theta$ term which has a period of one-half that of the sine or cosine – has an angle-average of zero over the top or bottom half of the solenoid (Δθ = π) at any radius. This is consistent with recent AB experiments by Becker and Batelaan [8] showing the absence of a time delay for the longitudinal propagation of an electron.

The dispersive (i.e., velocity-dependent) transverse force $F_y$ given by Eq. (25), on the other hand, has a $\cos^2\theta$ term which angle-averages to a non-zero value of 0.5, consistent in a general sense with the results of Becker *et al*. showing an asymmetry in the AB wavefunction envelope [1]. For a magnetic field pointing out of the page in the +*z* direction in Fig. 1, Fig. 2 shows that the velocity- and flux-dependent transverse force $F_y$ on a *negatively*-charged particle such as an electron is predicted to be in the +*y* direction, with an angle-averaged magnitude in the steady-state given by one-half the magnitude of Eq. (25). Due to the dot product in Eq. (10), reversing the magnetic-field direction reverses the direction of the force, as has also been measured by Becker *et al*. [1].

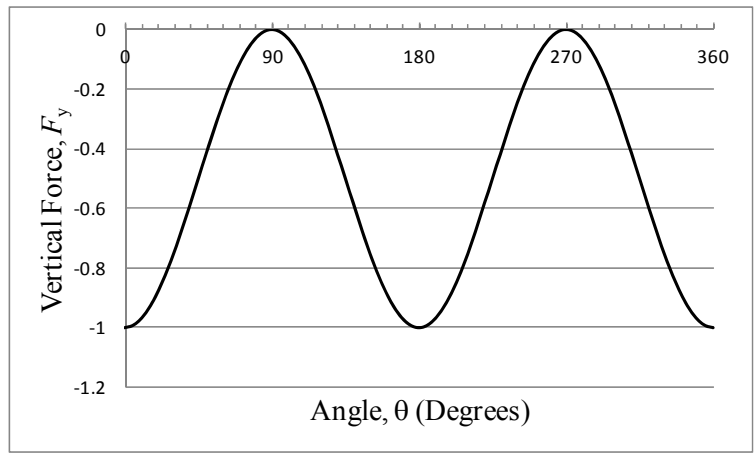

Fig. 2 – Angular dependence of the $-\cos^2\theta$ term for the vertical force component given by Eq. (41) for a positively-charged particle. For a particle moving from left-to-right in Fig. 1, the upper half of the solenoid starts at θ = 180 degrees at the leading edge and continues to θ = 0 degrees at the trailing edge; the lower half of the solenoid corresponds to θ = 180 to 360 degrees.

Potential-flow solutions to Laplace's equation also illustrate the difference in velocity across the upper and lower halves of the solenoid required for the fringe shift in the AB effect [17]. This is shown in Fig. 3, where the circulation term in Eq. (21) results in an asymmetry in the velocity distribution. This determines the difference in de Broglie wavelengths, and thus the phase difference between the upper and lower halves. Figure 3 is also consistent with the results of Becker and Batelaan [8] showing the absence of a time delay for longitudinal propagation. Physically, the acceleration and deceleration over the upper and lower halves of the solenoid balance over the electron paths for the Aharonov-Bohm geometry, with no change in the longitudinal propagation time of the electron's centroid.

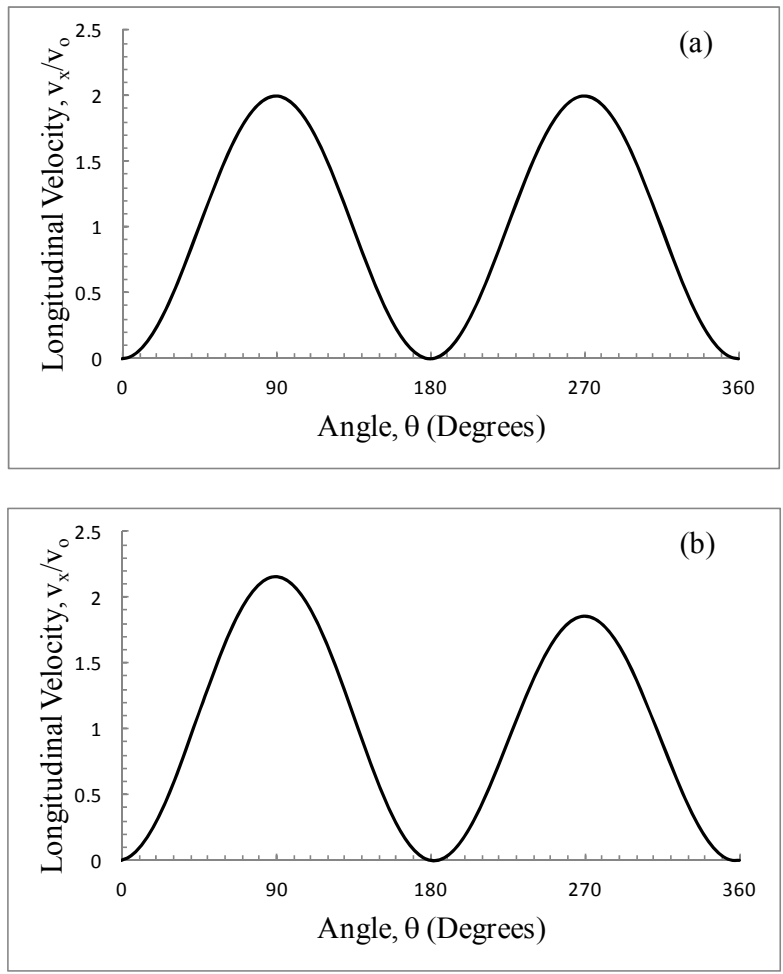

Fig. 3 – Angular dependence of the normalized x-component of the electron velocity at $r = R$ where $v_r = 0$ and $v_x = -v_\theta \sin\theta$. Shown is a comparison of (a) the velocity symmetry across the top half ($\theta = 180 \to 0$ degrees) and bottom half ($\theta = 180 \to 360$ degrees) of the solenoid when $\Phi_B = 0$; and (b) the asymmetry when $\Phi_B$ is not zero. In practice, the difference in normalized velocities will be on the order of $2eA_\theta/mv_o \approx 1$ part in $10^6$ for $v_o = 0.6 \times 10^8$ m/s ($E_o = 10$ keV).

The asymmetric velocity distribution shown in Fig. 3(b) can be used to obtain the expression $\Delta\phi_{AB} = e\Phi_B/\hbar$ for the AB phase shift. The phase of the electron de Broglie wave is given by $\phi = \mathbf{k} \cdot \mathbf{s}$, from which we find the phase difference $\Delta\phi$ between the top and bottom halves of the solenoid

$$\Delta\phi = \mathbf{s} \cdot \Delta\mathbf{k} = r \cdot \Delta k_r + r\theta_t \cdot \Delta k_\theta = R\theta_t \cdot \Delta k_\theta \tag{26}$$

for a path difference $\Delta s = 0$, a traversed angle $\theta_t$, and a difference in wavenumber $\Delta k_\theta$ based on the difference in azimuthal velocity $v_\theta$ at $r = R$ where $v_r = 0$. Writing out the wavenumber for the electron for $\psi_1$ and $\psi_2$ at the top and bottom of the solenoid, we have

$$k_{\theta 1} = \frac{2\pi}{\lambda_1} = \frac{m|v_{\theta 1}|}{\hbar} \quad \text{and} \quad k_{\theta 2} = \frac{2\pi}{\lambda_2} = \frac{m|v_{\theta 2}|}{\hbar} \tag{27}$$

where Eqns. (21) and (22) give

$$|v_{\theta 1}| = \left| -2v_o \sin\theta_1 - \frac{e}{m}\frac{\Phi_B}{2\pi R} \right| = \left| 2v_o \sin\theta_1 + \frac{e}{m}\frac{\Phi_B}{2\pi R} \right| \tag{28}$$

and

$$|v_{\theta 2}| = \left| -2v_o \sin\theta_2 - \frac{e}{m}\frac{\Phi_B}{2\pi R} \right| = \left| 2v_o \sin\theta_2 + \frac{e}{m}\frac{\Phi_B}{2\pi R} \right| \tag{29}$$

Substituting these results in Eq. (26), we obtain

$$\Delta\phi = \pi R \cdot \Delta k_\theta = R\theta_t \frac{m}{\hbar}\left[|v_{\theta 1}| - |v_{\theta 2}|\right] \quad (30)$$

Starting at the leading edge of the solenoid where $\theta_1 = \theta_2 = \pi$ rads, and propagating to the trailing edge where $\theta_1 = 0$ and $\theta_2 = 2\pi$ radians, we evaluate Eq. (30) numerically, obtaining a constant value for the difference in azimuthal velocities everywhere except at the points for the leading and trailing edges (where a difference cannot be defined). That constant depends on the velocity-circulation term in Eq. (21)

$$|v_{\theta 1}| - |v_{\theta 2}| = \frac{\Gamma_v}{\pi R} = \frac{e}{m}\frac{\Phi_B}{\pi R} \quad (31)$$

which, when combined with Eq. (30) and $\theta_t = \pi$ rads for $\psi_1$ and $\psi_2$, gives us the well-known non-dispersive expression $\Delta\phi_{AB} = e\Phi_B/\hbar$ for the AB phase shift. The magnetic flux $\Phi_B$ in the solenoid – and the resulting vector potential **A** in the field-free region outside the solenoid – thus determine both the phase shift (via the tangential component of the velocity) and the lateral envelope-displacement force (via the radial component of the velocity) in the AB effect.

V. SUMMARY and CONCLUSIONS

Summarizing, we use a second-order Taylor-series expansion of the vector potential in a Lagrangian analysis to identify an electromagnetic force in the AB effect. This force is a result of a charged particle moving through second-order spatial variations in the vector potential, with the acceleration and deceleration balancing over the propagation-symmetric path length for the Aharonov-Bohm geometry, thus giving no change in longitudinal propagation time in comparison with free space. As seen in Eq. (25), however, a gauge-invariant, velocity- and flux-dependent lateral force is predicted to displace an electron transverse to its longitudinal propagation direction.

This force can also be seen as a result of conservation of canonical angular momentum in Eq. (12). Any increases in mechanical angular momentum $mr^2\dot{\theta}$ are therefore due to decreases in field angular momentum $erA_\theta(r)$ associated with the second-order vector potential term.

Our results are consistent with all known AB experiments – including phase shift, lateral force, and absence of longitudinal force – and do not illustrate any conflict between "phase" and "force" arguments as to a possible AB mechanism. Instead, "phase" and "force" are both a consequence of the non-linear vector potential term in the region outside the solenoid.

Endnotes and References